\documentclass[preprint,12pt]{elsarticle}

\usepackage[T1]{fontenc} 

\newcommand{\Z}{\mathbb Z}

\usepackage{color}
\usepackage{url}

\newcommand{\BMSt}{\ensuremath{\mathfrak{bms}_3\ }}

\usepackage{amsmath, amsfonts, amssymb}
\usepackage{textgreek, stackengine, mathrsfs}
\usepackage{hyperref}
\usepackage[toc,page]{appendix}

\usepackage{amssymb}

\usepackage{color}
\usepackage[normalem]{ulem}

\numberwithin{equation}{section}

\newcommand{\dk}{d\widetilde{k}\,}

\newcommand{\ak}{a(\vec{k},t)}
\newcommand{\akd}{a^*(\vec{k},t)}

\begin{document}
	
\begin{frontmatter}

\title{A canonical realization of the Weyl BMS symmetry\tnoteref{prepn}}

\author{Carles Batlle\corref{cor1}\fnref{a}}
\ead{carles.batlle@upc.edu}
\author{V\'ictor Campello\fnref{b}}
\ead{vicmancr@gmail.com}
\author{Joaquim Gomis\fnref{c}}
\ead{joaquim.gomis@ub.edu}

\address[a]{Departament de Matem\`atiques and IOC, 
	Universitat Polit\`ecnica de Catalunya\\
	EPSEVG, Av. V. Balaguer 1, E-08800 Vilanova i la Geltr\'u, Spain}
\address[b]{Departament de Matem\`atiques i Inform\`atica, Universitat de Barcelona, Gran Via de les Corts Catalanes 585, 
	E-08007, Barcelona, Spain}
\address[c]{Departament de F\'isica Qu\`{a}ntica i Astrof\'isica and Institut de Ci\`{e}ncies del Cosmos (ICCUB), Universitat de Barcelona, Mart\'i i Franqu\`{e}s 1, E-08028 Barcelona, Spain
}

\tnotetext[prepn]{ICCUB preprint number: {\rm ICCUB-20-017}}	
\cortext[cor1]{Corresponding author}

\begin{abstract}
We construct a free field realization of an extension of the BMS algebra in $2+1$ dimensional space-time. Besides the supertranslations and superrotations, the extension contains an infinite set of superdilatations.
 We also comment 
  the difficulties that appear when trying to extend the algebra to that of the full
 conformal group.
\end{abstract}

\begin{keyword}
BMS symmetry, Conformal symmetry
\end{keyword}

\end{frontmatter}

\flushbottom

\section{Introduction}

There is a renewed interest in the BMS group \cite{Bondi:1962px,Sachs:1962wk}. One of the interest is to deduce Weinberg's soft graviton theorems \cite{Weinberg:1965nx} as the Ward identities of 
BMS supertranslations \cite{He:2014laa,He:2014cra,Campiglia:2015qka,Kapec:2015ena}. In this case the BMS symmetry is spontaneously broken. 
A pedagogical overview of the recent role of BMS symmetries is presented in \cite{Strominger:2017zoo}. 

In the case of three dimensions the BMS algebra has been studied
\cite{Ashtekar:1996cd,Barnich:2006av} with supertranslations and superrotations \cite{Barnich:2010eb}. A canonical realization of the \BMSt algebra  with supertranslations and superrotations  associated to a free 
Klein-Gordon (KG) field in $2+1$ dimensions, for both  massive and massless fields, was studied in \cite{Batlle:2017llu}. The canonical approach to BMS symmetry was introduced in \cite{Longhi:1997zt}.

At the quantum level, the Hilbert space of one-particle states supports a unitary irreducible representation of the Poincar\'e group, and at the same time a unitary reducible representation of the BMS$_3$ group. In contrast with the gravitational approach, our canonical realization of the supertranslations symmetry is not spontaneously broken. 

In this paper we reconsider the canonical realization associated to a 
massless KG field in terms of the Fourier modes.  As is well known the massless KG Lagrangian has an enlarged Poincar\'e symmetry
with dilatations and special conformal transformations. 

Knowing the fact that the Poincar\'e symmetry can be extended to a Poincar\'e
BMS symmetry,
it is natural to look for a possible infinite
dimensional extension of dilatations and special conformal
transformations.

We perform the analysis in the simple case of $d=3$. We will see that
that there is a natural generalization of dilatations to superdilatations that together with the supertranslations and superrotations close in an algebra that we call Weyl BMS algebra, as it is a generalization of the standard Weyl (Poincar\'e plus dilatations) algebra.\cite{Weyl1929b,NIEDERLE1974183,doi:10.1063/1.522756} We have not been able to construct a generalization
of special conformal transformations that close with supertranslations, 
superrotations and superdilatations.

The organization of the note is as follows. In Section \ref{BMSMKG} we construct the generators of the conformal algebra in terms of the Fourier modes of a massless KG field in arbitrary spacetime dimensions. In Section \ref{BMSMKG} we specialize to the case $d=3$ and give the expression of the BMS supertranslations and superrotations for the same field. Finally, in Section \ref{SDMKG} we introduce the superdilatations and explain the problems that appear when one considers the full BMS plus conformal algebra.

{\bf Note added.} When this paper was being completed a paper 
\cite{Donnay:2020fof} has appeared that constructs a generalization of the BMS symmetry by studying asymptotic symmetries of a gravitational theory. Their algebra, that includes superdilatations, agrees with our canonical results. A more general version of this algebra has also been found in \cite{Adami:2020ugu}, where general boundary symmetries in $d=2$ and $d=3$ are studied.

\section{Conformal algebra realization in terms of a  massless free Klein-Gordon field}
\label{CAMKG}
In this section we will construct  a realization of the conformal group 
in terms of the Fourier modes of a free massless scalar field.
Let us consider a massless scalar field theory with 
Lagrangian\footnote{The signature is $\eta = (-,+,\dots,+)$ for a $d$-dimensional spacetime.}
\begin{equation} \label{eq:lagrangian}
	\mathcal{L} = \dfrac{1}{2} \partial_\mu \phi \partial^\mu \phi, 
\end{equation}
The solution of the equations of motion can be written in terms of Fourier modes as
\begin{align}
	\phi(t, \vec{x}) &= \int \dk \left( \ak e^{i\vec{k} \cdot \vec{x}} + \akd e^{-i\vec{k} \cdot \vec{x}} \right),\nonumber\\ 
	& \dk = \dfrac{d^{d-1}k}{(2\pi)^{(d-1)}(2\omega)}, \quad \omega = k^0 = \sqrt{\vec{k}\cdot\vec{k}}.\label{solphi}
\end{align}

Following standard procedures, see for example \cite{Longhi:1997zt,Longhi:1998yd,Qualls:2015qjb}, one can construct the generators of the conformal algebra (Poincaré, dilatations and special conformal transformations) as integrals  over the space of local densities depending on the field and  their space-time derivatives.
In terms of the Fourier modes $\ak$, $\akd$  defined in (\ref{solphi}) they are given by
\begin{align}
	& P^\mu = \int \dk \akd k^\mu \ak, \quad k^\mu = (\omega, \vec{k}), \label{eq:momenta} \\
	& M^{0j} = tP^j - i\int \dk \akd \omega \dfrac{\partial}{\partial k_j} \ak, \qquad M^{j0} = - M^{0j}, \label{eq:boost} \\
	& M^{ij} = - i\int \dk \akd \left( k^i \dfrac{\partial}{\partial k_j} - k^j \dfrac{\partial}{\partial k_i} \right) \ak, \qquad M^{ji} = - M^{ij}, \label{eq:rotation} \\
	& D = -tP^0 + i \int \dk \akd \left( k^j \dfrac{\partial}{\partial k^j} + \dfrac{d-2}{2} \right) \ak, \label{eq:dilatation} \\
	& K^0 = - t^2 P^0 - \int \dk \akd \left[ -\left( \omega \dfrac{\partial}{\partial k_i} + 2it k^i \right)  \dfrac{\partial}{\partial k^i} - it(d-2) \right] \ak, \label{eq:special_conformal_time} \\
	& K^j = t^2 P^j - \int \dk \akd \left[ - \left( k^j \dfrac{\partial}{\partial k_i} - 2k^i \dfrac{\partial}{\partial k_j} \right) \dfrac{\partial}{\partial k^i} + (2it\omega + (d-2)) \dfrac{\partial}{\partial k_j} \right] \ak. \label{eq:special_conformal_space}
\end{align}
The equal-time canonical Poisson brackets between $\phi(t,\vec{x})$ and $\pi(t,\vec{x})=\dot{\phi}(t,\vec{x})$ lead to the following Poisson brackets for the 
  Fourier modes 
\begin{equation}
\{ \ak, \akd \} = -i (2\pi)^{(d-1)} (2\omega) \delta^{d-1}(\vec{k}-\vec{q}).
\label{poissonladder}
\end{equation}
Using (\ref{poissonladder}) one can compute the brackets between the generators of the conformal algebra
\begin{align}
	& \{ D, P^\mu \} = P^\mu,\label{eq:conf_algebra:1} \\
	& \{ D, K^\mu \} = -K^\mu, \label{eq:conf_algebra:2} \\
	& \{ K^\mu, P^\nu \} = 2 (\eta^{\mu\nu} D + M^{\mu\nu}),\label{eq:conf_algebra:3} \\
	& \{ K^\mu, M^{\nu\sigma} \} = \eta^{\mu\sigma} K^\nu - \eta^{\mu\nu} K^\sigma , \label{eq:conf_algebra:4} \\
	& \{ P^\mu, M^{\nu\sigma} \} = \eta^{\mu\sigma} P^\nu - \eta^{\mu\nu} P^\sigma, \label{eq:conf_algebra:5}\\
	& \{ M^{\mu\nu}, M^{\sigma\rho} \} = \eta^{\mu\sigma} M^{\nu\rho} + \eta^{\nu\rho} M^{\mu\sigma} - \eta^{\mu\rho} M^{\nu\sigma} - \eta^{\nu\sigma} M^{\mu\rho}, \label{eq:conf_algebra:6}
\end{align}
with all the remaining brackets equal to zero.
The Poisson brackets are related to the commutators of the differential operators in $\vec{k}$ by means of
\begin{equation}
	\left\{P, Q\right\} = -i \int \dk \akd [\hat{P}, \hat{Q}] \ak,
	\label{PB-C}
\end{equation}
where $\hat{P}$, $\hat{Q}$ are the operators appearing in the integrals of conserved charges $P$ and $Q$, respectively, at $t=0$. Explicitly,
\begin{align}\label{diffoperators}
\hat{P}^\mu &= k^\mu,\\
\hat{M}^{0j} &= -i \omega\dfrac{\partial}{\partial k_j},\\
\hat{M}^{ij} &= -i \left( k^i \dfrac{\partial}{\partial k_j} - k^j \dfrac{\partial}{\partial k_i} \right),\\
\hat{D} &= i \left( k^j \dfrac{\partial}{\partial k^j} + \dfrac{d-2}{2} \right),\label{opD}\\
\hat{K}^0 &= \omega \dfrac{\partial^2}{\partial k_i^2},\label{opK0}\\
\hat{K}^j &= \left( k^j \dfrac{\partial}{\partial k_i} - 2k^i \dfrac{\partial}{\partial k_j} \right) \dfrac{\partial}{\partial k^i} -  (d-2) \dfrac{\partial}{\partial k_j}.\label{opKi}
\end{align}

\section{Canonical realization of the BMS algebra}
\label{BMSMKG}
Following the ideas in \cite{Longhi:1997zt}, a canonical
realization of the full BMS algebra \cite{Barnich:2011ct} (supertranslations and superrotations) in terms of 
Fourier modes of
a massless Klein-Gordon field in $2+1$ spacetime was constructed in \cite{Batlle:2017llu}. Using the notation of the previous section, the generators of supertranslations and superrotations are given  respectively by
\begin{align}
\mathcal{P}_\ell &= \int \dk \akd \hat{\mathcal{P}}_\ell \ak, \ m\in\Z,\\
\mathcal{R}_m &= \int \dk \akd \hat{\mathcal{R}}_m \ak, \ m\in\Z,
\end{align}
with
\begin{align}
\hat{\mathcal{P}}_\ell
 & = \omega_\ell = \omega^{1-\ell} (k^1 + i k^2)^\ell,\quad \omega=\sqrt{k_1^2+k_2^2}, \\
\hat{\mathcal{R}}_m & = \dfrac{1}{\omega} \omega_m \left( (k^2 + imk^1) \dfrac{\partial}{\partial k^1} - (k^1 - imk^2) \dfrac{\partial}{\partial k^2} \right).
\end{align}
These differential operators obey the BMS algebra
\begin{align}
[ \hat{\mathcal{P}}_m, \hat{\mathcal{P}}_\ell ] &=0,\\
[ \hat{\mathcal{R}}_m, \hat{\mathcal{P}}_\ell ] &= i(m-\ell) \hat{\mathcal{P}}_{m+\ell}, \\ 
[\hat{\mathcal{R}}_m, \hat{\mathcal{R}}_n ] &= i(m-n) \hat{\mathcal{R}}_{m+n}.
\end{align}

For $m=0,\pm 1$ one obtains a $6$-dimensional closed algebra   which corresponds to Poincar\'{e} in $2+1$. Corresponding expressions for the massive Klein-Gordon field are also presented in  \cite{Batlle:2017llu}, but we will not discuss them  here since we are interested in extending this algebra with conformal generators.

\section{Superdilatations}
\label{SDMKG}
 We have seen in  Section \ref{CAMKG} that there is a 
canonical realization of the conformal algebra in terms of a massless Klein-Gordon,  and in Section \ref{BMSMKG}, in the case of $d=3$, that there is a BMS extension of the Poincar\'{e} algebra. It is therefore natural to look, in this simple case, for an extension of BMS Poincar\'{e}  that also includes dilatations and  special conformal
transformations.

The differential operators  (\ref{opD}),  (\ref{opK0}), (\ref{opKi}), appearing in the canonical realization of the conformal group   are, for $d=3$,  
\begin{align}
\hat{D} &= i \left( k^j \dfrac{\partial}{\partial k^j} + \dfrac{1}{2} \right),\\
\hat{K}^0 &= \omega \dfrac{\partial^2}{\partial k_i^2},\\
\hat{K}^j &= \left( k^j \dfrac{\partial}{\partial k_i} - 2k^i \dfrac{\partial}{\partial k_j} \right) \dfrac{\partial}{\partial k^i} -  \dfrac{\partial}{\partial k_j}.
\end{align}

 For the dilatations the commutations relations with the Poincar\'e BMS
generators are 
\begin{align}
[ \hat{D}, \hat{\mathcal{P}}_\ell] &= i\hat{\mathcal{P}}_\ell, \\
 [ \hat{D}, \hat{\mathcal{R}}_m ] &= 0, 
\end{align}
but when acting  with the special conformal transformations on the supertranslations one gets  new operators not present in the algebra
(\ref{diffoperators})
\begin{align}
	 -i[ \hat{K^1}, \hat{P_\ell} ] &=  -(1-\ell)\ell \dfrac{1}{\omega} \omega_{\ell+1} \hat{D} + (1+\ell)\ell \dfrac{1}{\omega} \omega_{\ell-1} \hat{D} + (1-\ell) \hat{\mathcal{R}}_{\ell+1} - (1+\ell) \hat{\mathcal{R}}_{\ell-1} ,\\
	 [ \hat{K^2}, \hat{P_\ell} ] &=  -(1-\ell)\ell \dfrac{1}{\omega} \omega_{\ell+1} \hat{D} - (1+\ell)\ell \dfrac{1}{\omega} \omega_{\ell-1} \hat{D} + (1-\ell) \hat{\mathcal{R}}_{\ell+1} + (1+\ell) \hat{\mathcal{R}}_{\ell-1} ,
\end{align}
This algebra can be closed if we introduce
an infinite family of superdilatations, given by
\begin{equation}
\hat{D}_\ell = \dfrac{1}{\omega} \omega_\ell \hat{D},
\end{equation}
so that
\begin{align}	
 -i[ \hat{K}^1, \hat{P_\ell} ] &=  -(1-\ell)\ell \hat{D}_{\ell+1} + (1+\ell)\ell \hat{D}_{\ell-1} + (1-\ell) \hat{\mathcal{R}}_{\ell+1} - (1+\ell) \hat{\mathcal{R}}_{\ell-1} ,\\
 [ \hat{K}^2, \hat{P_\ell} ] &=  -(1-\ell)\ell \hat{D}_{\ell+1} - (1+\ell)\ell \hat{D}_{\ell-1} + (1-\ell) \hat{\mathcal{R}}_{\ell+1} + (1+\ell) \hat{\mathcal{R}}_{\ell-1}.
\end{align}
Using $\hat{D}_\ell$, the commutator of $\hat{K}^0$ with the supertranslations also closes,
\begin{equation}
[ \hat{K}^0, \hat{P}_\ell ] =  -2i \left( (1-\ell^2) \hat{D}_\ell + \ell \hat{\mathcal{R}}_\ell \right).
\end{equation}
These commutators yield Poisson brackets, via (\ref{PB-C}), which generalize 
(\ref{eq:conf_algebra:3}) when the supertranslations are considered.

It remains to study the action of the special conformal transformations on the superrotations. Using $\hat{K}^{\pm}$ instead of $\hat{K}^{1,2}$, given by 
\begin{equation}
\hat{K}^{\pm} = \dfrac{1}{2} \left(\hat{K}^{1} \pm i \hat{K}^{2}\right)
\end{equation}
one gets
\begin{align}
& \left[ \hat{K}^+, \hat{\mathcal{R}}_\ell \right] = -i\dfrac{1}{2} \ell (1-\ell)\dfrac{1}{\omega} \omega_{\ell+1} \hat{K}^0 + i(1-\ell^2)\dfrac{1}{\omega} \omega_\ell \hat{K}^+ + i\dfrac{1}{2} \ell (1-\ell^2) \dfrac{1}{\omega^2} \omega_{\ell+1} k^i \partial_i, \label{KpR} \\
& \left[ \hat{K}^-, \hat{\mathcal{R}}_\ell \right] = -i\dfrac{1}{2} \ell (1+\ell)\dfrac{1}{\omega} \omega_{\ell-1} \hat{K}^0 - i(1-\ell^2) \dfrac{1}{\omega} \omega_\ell \hat{K}^- + i\dfrac{1}{2} \ell (1-\ell^2) \dfrac{1}{\omega^2} \omega_{\ell-1} k^i \partial_i, \label{KmR} \\
& \left[ \hat{K}^0, \hat{\mathcal{R}}_\ell \right] = -i\ell (1+\ell) \dfrac{1}{\omega} \omega_{\ell-1} \hat{K}^+ - i\ell (1-\ell) \dfrac{1}{\omega} \omega_{\ell+1} \hat{K}^- + i\ell(1-\ell^2) \dfrac{1}{\omega^2} \omega_\ell k^i \partial_i,\label{K0R}
\end{align}
and new operators appear. One can be tempted, as we did for superdilatations, to introduce a family of superspecial conformal transformations in order to take into account the first and second terms in the right-hand side of the above commutators. However, the last term cannot be absorbed, unless one introduces yet another, completely new family of operators, and we have not succeed in obtaining a close algebra which includes both the supertranslations and superrotations and the special conformal transformations.   Notice also that one cannot just drop the superrotations, since they appear in the commutator between special conformal transformations and supertranslations. The extra terms in (\ref{KpR})-(\ref{K0R}) disappear for $\ell=0,\pm1$ as it must be, since then we have a part of the conformal algebra.

Leaving aside the special conformal transformations, the superdilatations, together with the supertranslations and the superrotations, form a closed algebra, which we call a Weyl BMS algebra, given by
\begin{align}
[ \hat{\mathcal{P}}_m, \hat{\mathcal{P}}_\ell ] &=0,\\
[ \hat{\mathcal{R}}_m, \hat{\mathcal{P}}_\ell ] &= i(m-\ell) \hat{\mathcal{P}}_{m+\ell}, \\ 
[\hat{\mathcal{R}}_m, \hat{\mathcal{R}}_n ] &= i(m-n) \hat{\mathcal{R}}_{m+n},\\
[ \hat{D}_\ell, \hat{\mathcal{P}}_m ] &= i \hat{\mathcal{P}}_{\ell+m},\\
[\hat{D}_\ell, \hat{D}_m] &= 0, \\ 
[\hat{D}_\ell, \hat{\mathcal{R}}_m] &= i \ell \hat{D}_{\ell+m}.
\end{align}

\section{Conclusions}
We have obtained a family of operators, the superdilatations, that appear naturally when considering the action of the special conformal transformations on the supertranslations. The set of supertranslations, superrotations and superdilatations form a closed infinite dimensional algebra which can be considered a BMS extension of the ordinary Weyl algebra (Poincar\'e plus dilatations). Trying to include the special conformal transformations leads to the appearance of an infinite tower of new kinds of operators. A more detailed description of the difficulties encountered when trying to close the algebra in our approach will be reported in an extended work.

\section*{Acknowledgements}
We would like to thank Jos\'e F. Cari\~{n}ena for his comments about the bibliography of the Weyl group, and Hamed Adami for pointing out the relation of our results with those in  \cite{Adami:2020ugu}.
The work of CB has been
partially supported by  {Generalitat de Catalunya} through project   {SGR2017 872}, and by 
 {MINECO}
 {RTI2018-096001-B-C32}.
JG has been supported in part by
 {MINECO}
 {FPA2016-76005-C2-1-P}, {PID2019-105614GB-C21},
and by the Spanish government (MINECO/FEDER) under
project  {CEX2019-000918-M} of {Unidad de Excelencia Mar\'{i}a de Maeztu}

\bibliography{bib.bib}{}
\bibliographystyle{elsarticle-num}

\end{document}